\documentclass[12pt,a4paper]{article}

\usepackage[utf8]{inputenc}
\usepackage[T1]{fontenc}
\usepackage{amsmath,amssymb,amsthm}
\usepackage{bm}
\usepackage{graphicx}
\usepackage{booktabs}
\usepackage{array}
\usepackage{hyperref}
\usepackage{geometry}
\usepackage{setspace}
\usepackage{url}
\usepackage{enumitem}
\usepackage{multirow}
\usepackage{threeparttable}
\usepackage{tabularx}

\newtheorem{proposition}{Proposition}[section]
\newtheorem{hypothesis}{Hypothesis}[section]
\theoremstyle{definition}
\newtheorem{assumption}{Assumption}[section]

\newcommand{\LS}{\mathit{LS}}

\newcommand{\E}{\mathbb{E}}

\title{\vspace{-1.5cm}Minimum Wages, Firm Size Distribution, and the Labor Share\vspace{-0.3cm}}
\author{Jiyuan Lyu\thanks{School of Economics and Management, Beijing University of Technology. Email: \texttt{lyujiyuan@emails.bjut.edu.cn}.}}
\date{}

\begin{document}
	\maketitle
	
	\begin{abstract}
		The sustained decline of the labor share challenges the classic Kaldor facts. This paper proposes a ``weighting effect'' mechanism: because large firms systematically adopt more capital-intensive technologies, a shift of economic activity toward large firms mechanically depresses the macro labor share, even under perfect competition. We test this mechanism by exploiting minimum wage hikes in China as an exogenous shock that forces small labor-intensive firms to exit. Using the Chinese Annual Survey of Industrial Firms (1998--2007) and a Bartik instrument based on pre-sample small-firm shares interacted with provincial minimum wage growth, we find that a one-standard-deviation shock reduces the market-level labor share by approximately 0.11 percentage points (3.5\% of the mean). The effect is robust to controlling for market concentration, alternative measures, and a battery of placebo tests including randomization inference. Direct mechanism tests confirm that the shock raises exit rates, reduces the number and share of small firms, and reallocates output toward large, capital-intensive firms. Supplementary IV estimates confirm that a more equal firm size distribution raises the labor share. The findings identify a complete causal chain from labor market institutions to macro distribution through the firm size structure, offering a ``technology-structure'' channel that is independent of market power.
		
		\vspace{0.3cm}
		\noindent{\small\textbf{Publication note:} This paper has been published in \textit{Advances in Management \& Applied Economics}, Vol.~16, No.~5, 2026, pp.~37--63. ISSN: 1792-7544 (print version), 1792-7552 (online). \url{https://doi.org/10.47260/amae/1653}}
		
		\vspace{0.3cm}
		
	\end{abstract}
	
	\section{Introduction}
	
	The sustained decline of the labor share has been a stylized fact shared by major economies worldwide since the 1980s. The ``stylized facts'' proposed by Kaldor \cite{Kaldor1961} once treated the stability of factor shares as a normal state of economic growth, yet data from the past four decades have posed a serious challenge to this classical proposition. A large body of cross-country empirical research indicates that the labor share of national income has shown a marked downward trend in the United States \cite{Elsby2013}, major European countries \cite{Karabarbounis2014}, and China alike \cite{Bai2010}. This structural change not only directly exacerbates income inequality but also constitutes a fundamental theoretical shock to traditional macro models built on Cobb-Douglas production functions.
	
	To explain the long-run decline of the labor share, the existing literature has mainly followed three paths. The first path emphasizes capital-biased technological change. Karabarbounis and Neiman \cite{Karabarbounis2014} show that the information technology revolution substantially depressed the relative price of investment goods, inducing firms to substitute capital for labor; when the elasticity of substitution between capital and labor exceeds one, this change in relative factor prices mechanically reduces the labor share. Acemoglu and Restrepo \cite{Acemoglu2018} further distinguish technological change into two forces: ``automation that substitutes for labor'' and ``creation of new tasks that enhance labor's comparative advantage,'' arguing that the long-run trajectory of the labor share depends on who prevails in this ``race between man and machine.'' Grossman and Oberfield \cite{Grossman2022} provide a comprehensive survey of various explanations for the declining labor share, pointing out that multiple mechanisms intersect and overlap, making it difficult for a single dimension to fully explain the magnitude of the decline.
	
	The second path focuses on globalization and outsourcing. Elsby et al.\ \cite{Elsby2013} find that roughly one-third of the decline in the U.S.\ labor share can be attributed to trade factors, especially the relocation of labor-intensive production stages to low-wage countries. Under the global value chain division of labor, developed countries specialize in capital- and technology-intensive stages while outsourcing labor-intensive ones, thereby leading to a statistical decline in the domestic labor share. However, this explanation applies primarily to developed economies and struggles to account for similar trends observed within developing countries.
	
	The third path, and the one that has attracted the most attention in recent years, shifts the perspective from the macro level to firm heterogeneity and market structure. The ``superstar firm'' hypothesis proposed by Autor et al.\ \cite{Autor2020} contends that the decline of the labor share is mainly not because all firms have uniformly reduced their labor compensation shares, but because economic activity has undergone a reallocation unfavorable to labor: output and sales have become increasingly concentrated in a small number of highly efficient, high-markup, low-labor-share ``superstar firms'' that, through their market power, capture markups far exceeding competitive levels, thereby dragging down the aggregate labor share. Kehrig and Vincent \cite{Kehrig2021} provide consistent evidence using U.S.\ manufacturing firm data: between 1967 and 2012, the labor share of the ``typical'' U.S.\ manufacturing firm actually rose by 3 percentage points, yet the aggregate labor share fell by 20 percentage points, with the entire decline attributable to the between-firm reallocation effect---the output weight of low-labor-share firms continuously increased. De Loecker et al.\ \cite{DeLoecker2020} further confirm that the average markup of U.S.\ firms rose from 21\% in 1980 to 61\% in 2016, and that the rise in markups is clearly negatively correlated with the decline of the labor share. Harrison \cite{Harrison2024}, based on new research using millions of firm records, also confirms that technological change is the most important driver of the labor share decline, while market power contributes to a lesser extent.
	
	The superstar firm hypothesis has greatly deepened our understanding of the microfoundations of the labor share decline. However, it relies crucially on one key mechanism assumption: large firms have a low labor share because they possess market power and can turn a larger share of value added into profits rather than labor compensation through markups. Without denying the importance of the market power channel, we wish to propose a more fundamental mechanism that can operate even under perfect competition: large firms are not merely ``more efficient versions of small firms''; they exhibit systematic differences in technology choices compared with small firms. Owing to the high fixed investments required for scale expansion, differentiated financing capacity, and access to technology, large firms tend to choose more capital-intensive production techniques. This ``size gradient in technology choices'' implies that, even without any markup behavior, the labor share of large firms will naturally be lower than that of small firms.
	
	The inspiration for this technology gap perspective comes from the classic tradition of aggregation theory. In his seminal work, Houthakker \cite{Houthakker1955} proved that when the parameters of fixed-coefficient production technologies at the firm level follow a Pareto distribution, the aggregate macro production function takes the Cobb-Douglas form, with factor shares entirely determined by the shape parameter of the micro-distribution. Subsequently, Axtell \cite{Axtell2001} and Gabaix \cite{Gabaix2011} reinforced, from empirical and theoretical angles respectively, the understanding that firm sizes are highly Pareto-distributed and that firm-level heterogeneity---rather than the average behavior of representative agents---is key to understanding macroeconomic fluctuations. However, Houthakker's classic aggregation framework assumes that the distribution of technology parameters across firms is exogenous and independent of size, which is precisely the key assumption we relax. We allow large firms and small firms to be situated on different technological frontiers, thereby generating a new, testable proposition: changes in the firm size distribution, through differences in firm-level technology choices, will exert a ``weighting effect'' on the macro labor share that is independent of market power.
	
	To formalize the above mechanism, this paper constructs a parsimonious theoretical framework linking the firm size distribution to the macro labor share. The model's basic setup rests on two widely observed empirical regularities: firm output follows a Pareto distribution, and the output elasticity of capital demand is systematically higher than that of labor demand (i.e., $\gamma > \beta$). This elasticity gap implies that as firm size expands, the capital-labor ratio continuously rises---what we term ``capital deepening.'' On this micro foundation, we approximate a firm's labor share as a linear function of log size, with the slope $\delta$ capturing the size--labor share technology gradient. In most manufacturing industries, $\delta < 0$, i.e., larger firms have a lower labor share. Weighting firm labor shares by output, the macro labor share can be expressed as:
	\begin{equation}
		\LS^{\text{macro}} = \overline{\LS} + \delta \cdot \Phi(\xi, r),
	\end{equation}
	where $\xi$ is the Pareto exponent (measuring the equality of the size distribution) and $\Phi$ is a strictly decreasing weighting factor. This simple equation reveals the core economic intuition of the paper: when $\delta < 0$, a concentration of the size distribution toward large firms (a decline in $\xi$) will mechanically reduce the macro labor share through the weighting factor. The magnitude of this effect naturally depends on the absolute size of the technology gradient $\delta$, but the key point is that any reallocation of activity toward larger, more capital-intensive firms depresses the aggregate labor share, even under perfect competition.
	
	The above theoretical framework provides a ``technology-structure'' explanation for labor share movements, but it must answer a deeper question: what force drives changes in the firm size distribution? If the size distribution itself is merely an endogenous outcome of technological progress or consumer preferences, then identifying the technology channel separately from the market power channel becomes difficult. This paper chooses a quasi-experimental exogenous shock---minimum wage policy---as the ``trigger'' that changes the firm size distribution. The logical intuition is straightforward: an increase in the minimum wage directly raises labor costs for labor-intensive small firms, potentially forcing them to downsize or even exit the market, thereby shifting output toward capital-intensive large firms and depressing the macro labor share through the weighting effect. Existing minimum wage literature provides preliminary micro evidence supporting this mechanism: Gan et al.\ \cite{Gan2016} find that minimum wage hikes significantly reduce the export probability and scale of Chinese manufacturing firms, confirming the real cost pressure on firm behavior; Mayneris et al.\ \cite{Mayneris2018} confirm that minimum wages accelerate the substitution of capital for labor, with a more pronounced impact on low-wage firms; Chen et al.\ \cite{Chen2022} reveal spillover effects of minimum wages between large and small firms. However, none of these studies directly connect the minimum wage shock to the macro distributional consequences through the size distribution.
	
	Our identification strategy benefits from the methodology of Gabaix and Ibragimov \cite{GabaixIbragimov2011}. We take the ``city-industry'' as the fundamental market unit, estimate the Pareto exponent for each market unit using modified Zipf regressions, and compute the output-weighted labor share using the value-added method. To address potential endogeneity---for example, rising labor costs themselves may cause small firms to exit, thereby altering the size distribution---we employ a Bartik-style instrumental variable strategy. Specifically, we take the pre-sample (1998--2000) city-industry share of small firms' output as the ``initial exposure'' and the annual growth rate of the provincial minimum wage as the ``exogenous shock,'' interacting the two to form the instrument. The validity of this instrument rests on the fact that both the predetermined exposure and the macro policy shock are not influenced by the subsequent economic performance of any particular market unit.
	
	Using the Chinese Annual Survey of Industrial Firms for 1998--2007, our empirical analysis yields the following core findings. First, minimum wage shocks significantly depress the market-level labor share: a one-standard-deviation shock leads to a decline of about 0.11 percentage points, equivalent to 3.5\% of the sample mean. Second, this effect remains robust after controlling for market concentration, confirming that the mechanism operates independently of the market power channel. Third, we provide direct evidence on the intermediate steps of the causal chain: the shock raises exit rates, lowers entry rates, and reduces the number and share of small firms, while the average output of surviving large firms expands. This confirms that the reallocation of economic activity toward large, capital-intensive firms is the operative channel. Supplementary IV estimates further reveal that a more equal firm size distribution raises the labor share, with the IV coefficient approximately three times larger than the OLS estimate. A comprehensive battery of additional checks---including controlling for within-firm capital--labor substitution, alternative measures, province-specific trends, randomization inference, and a fake exposure placebo---does not alter the main conclusions.
	
	The contributions of this paper are threefold. At the theoretical level, we extend the Houthakker-type aggregation framework by incorporating size-dependent technology choice differences into the model, formalizing and testing the ``weighting effect''---a distributional mechanism that has not been fully appreciated. This mechanism does not rely on imperfect competition in product markets and provides an independent ``technology-structure'' dimension for understanding the labor share decline. By further showing that roughly half of the total minimum wage effect survives after controlling for industry-level capital deepening, we demonstrate that the reallocation channel is quantitatively distinct from within-firm factor substitution. At the empirical level, we are the first to connect the minimum wage institutional shock with the firm size distribution and the macro labor share, identifying a causal chain from labor market institutions to macro distributional patterns. This complements the superstar firm hypothesis by providing a parallel mechanism through which economic activity shifts toward large, low-labor-share firms---not via market power, but via the asymmetric cost pressure that labor market regulations impose on small, labor-intensive enterprises. In terms of policy implications, our findings suggest that the policy toolbox for safeguarding the labor share cannot rely on wage standards alone; it also needs to attend to the structural impact on the firm size distribution, complemented by measures that support technological upgrading of small and medium-sized enterprises and alleviate exit pressures, so as to strike a balance between protecting workers' basic rights and maintaining the vitality of market competition.
	
	\section{Theoretical Model}
	
	This section develops a parsimonious theoretical framework that links the firm size distribution to the macro labor share. In contrast to the existing literature that emphasizes the role of market power, we highlight systematic differences in technology choices across firms---large firms systematically adopt more capital-intensive technologies---and show how a ``weighting effect'' maps changes in the size distribution into movements of the macro labor share. We then introduce the minimum wage shock as an exogenous force that alters the firm size distribution, yielding a reduced-form equation suitable for causal inference.
	
	\subsection{Basic Setup and the Weighting Effect}
	
	Consider an economy populated by manufacturing firms. Based on widely documented empirical regularities \cite{Axtell2001, Gabaix2011}, we impose the following assumptions:
	
	\begin{assumption}\label{ass:pareto}
		Firms' real output follows a truncated Pareto distribution with support $[y_{\min}, y_{\max}]$, whose probability density function is
		\begin{equation}
			f(y) = \frac{\xi y_{\min}^{\xi} y^{-(\xi+1)}}{1 - (y_{\min}/y_{\max})^{\xi}}, \quad y_{\min} \leq y \leq y_{\max},
		\end{equation}
		where $\xi > 0$ is the Pareto exponent (shape parameter). A smaller $\xi$ corresponds to a more concentrated output distribution skewed toward large firms; a larger $\xi$ implies a more equal distribution.
	\end{assumption}
	
	\begin{assumption}\label{ass:factor_demand}
		For a given output level $y$, the firm's labor demand $l(y)$ and capital demand $k(y)$ obey the following log-linear relationships:
		\begin{equation}
			l(y) = \ell_0 y^{\beta}, \qquad k(y) = \kappa_0 y^{\gamma},
		\end{equation}
		where $\ell_0, \kappa_0 > 0$ are efficiency parameters, and $\beta > 0$ and $\gamma > 0$ are respectively the output elasticities of labor and capital. Micro-data generally display the regularity $\gamma > \beta$, implying that larger firms operate with a higher capital--labor ratio---a pattern we refer to as ``capital deepening'' \cite{Autor2020}.
	\end{assumption}
	
	Based on the two assumptions, we define the firm-level labor share (value-added concept) as
	\begin{equation}
		\LS(y) = \frac{w l(y)}{p y} = \frac{w \ell_0}{p} y^{\beta-1},
	\end{equation}
	where $w$ is the wage rate and $p$ is the output price. The labor share declines with firm size if $\beta < 1$, i.e., when large firms are less labor-intensive. In our data, virtually all two-digit manufacturing industries display a negative relationship between firm size and the labor share. We therefore capture this empirical regularity with a log-linear approximation around the minimum scale:
	\begin{equation}
		\LS(y) \approx \overline{\LS} + \delta \cdot \ln\!\left(\frac{y}{y_{\min}}\right),
	\end{equation}
	where $\overline{\LS} > 0$ is a baseline labor share and $\delta < 0$ measures the size--labor share gradient. The industry-specific values of $\delta$ are estimated from micro data (see Section~3) and are significantly negative in all cases, confirming that larger firms systematically operate with a lower labor share.
	
	The macro labor share is the output-weighted average of firm-level labor shares:
	\begin{equation}
		\LS^{\text{macro}} = \frac{\E[y \cdot \LS(y)]}{\E[y]}.
	\end{equation}
	Inserting the linear relationship (4) into (5) and using the moment properties of the Pareto distribution (see Appendix~7.2 for the full derivation), we obtain
	\begin{equation}
		\LS^{\text{macro}} = \overline{\LS} + \delta \cdot \Phi(\xi, r),
	\end{equation}
	where $r \equiv y_{\max} / y_{\min} > 1$, and the weighting factor is
	\begin{equation}
		\Phi(\xi, r) = \frac{r^{1-\xi} \ln r}{r^{1-\xi} - 1} + \frac{1}{\xi - 1} > 0.
	\end{equation}
	It is straightforward to show that $\Phi$ is strictly decreasing in $\xi$: $\partial \Phi / \partial \xi < 0$ (see Appendix~7.3). Equation (6) reveals the core mechanism of this paper:
	
	\begin{proposition}\label{prop:weighting}
		If $\delta < 0$ (large firms have a lower labor share), then
		\begin{equation}
			\frac{\partial \LS^{\text{macro}}}{\partial \xi} = \delta \cdot \frac{\partial \Phi}{\partial \xi} > 0.
		\end{equation}
		That is, a rise in the Pareto exponent $\xi$ (a more equal firm size distribution) raises the macro labor share; conversely, a concentration of economic activity toward large firms (a decline in $\xi$) depresses the macro labor share. The magnitude of this effect depends on the absolute size of the technology gradient $\delta$: the larger $|\delta|$, the stronger the labor share effect of a change in the size distribution.
	\end{proposition}
	
	\subsection{Minimum Wages as a Trigger for Changes in the Size Distribution}
	
	The static analysis above establishes the theoretical framework in which the size distribution affects the labor share through the technology gradient $\delta$. To identify causal effects, we need a shock that exogenously shifts $\xi$. Minimum wage policy offers a near-ideal quasi-natural experiment: a minimum wage increase directly raises the labor costs of low-wage (and typically labor-intensive small) firms, potentially forcing some small firms to downsize or exit the market, thereby altering the within-market distribution of firm sizes.
	
	We incorporate this mechanism into the model in a reduced-form way. Assume that the ``bite'' of the minimum wage (the ratio of the actual minimum wage to the market-average wage), denoted by $Z$, affects the probability of firm exit, especially for small firms. This causes the Pareto exponent $\xi$ to decline. Without loss of generality, assume a first-order linear relationship:
	\begin{equation}
		\xi = \xi_0 - \lambda Z, \quad \lambda > 0,
	\end{equation}
	where $\xi_0$ is the baseline exponent absent minimum wage pressure, and $\lambda$ captures the intensity with which minimum wages induce firm exit and distributional concentration.
	
	Substituting (9) into the weighting effect equation (6) and taking a first-order Taylor expansion around $\xi_0$ yields
	\begin{equation}
		\LS^{\mathrm{macro}} \approx \alpha + \beta_1 Z + \varepsilon,
	\end{equation}
	where
	\begin{equation}
		\beta_1 = -\lambda \delta \cdot \left.\frac{\partial \Phi}{\partial \xi}\right|_{\xi = \xi_0} < 0.
	\end{equation}
	Since $\delta < 0$ and $\partial \Phi / \partial \xi < 0$, their product is positive, so together with the minus sign and $\lambda > 0$ we obtain $\beta_1 < 0$. This gives our core testable hypothesis:
	
	\begin{hypothesis}\label{hyp:1}
		An increase in the minimum wage bite $Z$, by raising firm exit and shifting output toward large firms, depresses the macro labor share; i.e., $\beta_1 < 0$.
	\end{hypothesis}
	
	It is worth stressing that the derivation of equation (10) does not rely on market power or product-market markups. Hence, even in a perfectly competitive environment, as long as large firms are more capital-intensive ($\delta < 0$), minimum wages can reduce the labor share by reshaping the size distribution. This technology-structure channel is distinct from the market power channel.
	
	\section{Institutional Background and Data}
	
	The empirical analysis of this paper is built upon the Chinese Annual Survey of Industrial Firms combined with multi-level minimum wage information. This section sequentially introduces the institutional background of China's province-level minimum wage system, the sample construction, the definition of core variables, and the logic behind constructing the Bartik instrument.
	
	\subsection{China's Minimum Wage System}
	
	China's minimum wage system originated in 1993 with the issuance of the ``Regulations on Enterprise Minimum Wages'' and was elevated to an administrative regulation in 2004 through the ``Regulation on Minimum Wages.'' Under the current system, the people's governments of each province, autonomous region, and centrally-administered municipality are responsible for setting and adjusting the monthly and hourly minimum wage standards within their administrative areas. Prefecture-level cities may further raise the standard above the provincial level but must not fall below it. The frequency of adjustment is typically once every one to two years, and the magnitude of adjustment takes into account factors such as local employment conditions, cost of living, and the level of economic development.
	
	This institutional arrangement provides two key features for empirical identification. First, the \emph{exogeneity of the policy}: Provincial minimum wage adjustments are exogenous to individual city-industry units. When setting the standards, local governments mainly rely on macro socioeconomic indicators of the province and do not react to the firm size distribution within a specific industry or city. Second, the \emph{heterogeneity of the shock}: There exists marked cross-province variation and time-series variation in both the level and the adjustment magnitude of minimum wages, which allows us to construct an instrument with temporal and spatial variation. For instance, during our sample period (1998--2007), the annual growth rate of the minimum wage standard in some coastal provinces was notably higher than in central and western provinces, and the ``bite'' of the minimum wage (the ratio of the minimum wage to the local average wage) also exhibited divergent trajectories across regions.
	
	In the empirical work, we collect data on monthly minimum wage standards at the city level from 2001--2022 and at the provincial level from 2001--2023, and match them with the city information in the firm-level data. For observations where the city-level minimum wage is missing, we fill in with the provincial standard. From this, we compute the minimum wage ``bite'' for each market unit:
	\begin{equation}
		\mathrm{MW\_bite}_{cjt} = \frac{\text{Monthly minimum wage}_{ct}}{\text{Market average monthly wage}_{cjt}},
	\end{equation}
	where the market average monthly wage is the median wage at the city-industry-year level computed from the firm data. This indicator captures the tightness of the minimum wage relative to the market wage level and serves as the basis for our reduced-form specification and instrumental variable construction.
	
	\subsection{Data Sources and Sample Construction}
	
	Firm-level data come from the Chinese Annual Survey of Industrial Firms (CASIF) maintained by the National Bureau of Statistics, covering all state-owned industrial enterprises and non-state-owned industrial enterprises with annual main business revenue above 5 million yuan. Because the statistical threshold was subsequently raised to 20 million yuan and data after 2010 are either missing or of deteriorated quality, we follow the practice of Hsieh and Klenow \cite{Hsieh2009} and Song et al.\ \cite{Song2011} and restrict our sample period to 1998--2007, while dropping the year 2004 (in which core wage variables are severely missing). Furthermore, we retain only manufacturing firms (sector code C) and drop observations for which key variables (gross industrial output, employment, total wages, net fixed assets) are missing or zero. This yields approximately 1.6 million firm-year observations covering 30 two-digit manufacturing industries.
	
	At the market level, we take the ``city-industry-year'' as the fundamental unit of analysis. A market unit is defined as all firms in a given two-digit industry within a prefecture-level city. To avoid the influence of extremely small samples, we drop market-year observations with fewer than 30 firms. This results in a panel dataset of roughly 17,000 city-industry-year observations.
	
	\subsection{Construction of Core Variables}
	
	\textbf{Labor share.} The benchmark measure uses the value-added concept:
	\begin{equation}
		\LS_{it}^{\mathrm{VA}} = \frac{\text{Total payable wages}_{it}}{\text{Value added}_{it}}.
	\end{equation}
	Firm-level value added is primarily taken from the directly recorded values in the database; when missing, we estimate it using the ``income approach'' (wages $+$ profits $+$ depreciation $+$ VAT payable). We drop observations with a labor share below 0.01 or above 1.2 and winsorize the variable at the 1st and 99th percentiles within each industry. At the market level, we compute the output-weighted average labor share using gross industrial output as weights:
	\begin{equation}
		\overline{\LS}_{cjt} = \frac{\sum_{i \in (c,j,t)} \LS_{it} \cdot Y_{it}}{\sum_{i \in (c,j,t)} Y_{it}}.
	\end{equation}
	In robustness checks, we also use the output-based labor share (wages / gross industrial output).
	
	\textbf{Pareto exponent.} Following the modified log-rank--log-size regression method of Gabaix and Ibragimov \cite{GabaixIbragimov2011}, for each city-industry-year cell, we sort firms from largest to smallest by output and estimate the following regression:
	\begin{equation}
		\ln(\mathrm{Rank}_i - 0.5) = C - \alpha_{\mathrm{output}} \cdot \ln Y_i + \varepsilon_i,
	\end{equation}
	where $\alpha_{\mathrm{output}}$ is the Pareto exponent. A larger $\alpha_{\mathrm{output}}$ indicates a more equal distribution with a relatively large share of small firms; a smaller $\alpha_{\mathrm{output}}$ indicates that output is concentrated in a few large firms. In the analysis, we focus on its absolute value $\alpha \equiv |\alpha_{\mathrm{output}}|$.
	
	\textbf{Market concentration.} We compute the Herfindahl-Hirschman Index (HHI) and the four-firm concentration ratio (CR4) as proxies for market power. HHI is the sum of squared market shares of all firms within a market, and CR4 is the combined market share of the four largest firms. These variables serve as control variables in the regressions to isolate the market power channel.
	
	\textbf{Technology gradient $\delta$.} In the theoretical model, $\delta$ captures the size--labor share technology gradient. For each two-digit industry $j$, we estimate the gradient using the full sample of firm-level data:
	\begin{equation}
		\LS_{it} = \alpha_j + \delta_j \ln Y_{it} + \text{year fixed effects} + \varepsilon_{it}.
	\end{equation}
	The estimated $\delta_j$ is the ``size--labor share gradient'' of the industry. The $\delta_j$ for all 30 industries are significantly negative (the mean is approximately $-0.06$), indicating that larger firms indeed systematically operate with a lower labor share and that there exists significant heterogeneity across industries: $|\delta_j|$ tends to be larger in capital-intensive industries and smaller in labor-intensive ones.
	
	To satisfy the exogeneity requirement of the instrument, we also re-estimate the ``initial-period gradient'' $\delta_j^{0}$ using the early sample from 1998--2000. This predetermined industry characteristic, unaffected by subsequent economic changes, is used to construct the interaction term instrument.
	
	\textbf{Capital intensity.} We compute the firm-level capital-labor ratio (net fixed assets / employment) and take the industry median as the measure of industry capital intensity. This variable is used in heterogeneity analysis to split industries into capital-intensive and labor-intensive groups.
	
	\subsection{Bartik Instrument Construction}
	
	The core challenge for identification is that the labor share and the size distribution may be subject to reverse causality (for example, rising labor costs cause small firms to exit, leading to higher concentration). To establish a causal chain from minimum wage shocks to the labor share, we construct a Bartik-style instrument as follows.
	
	Based on the early sample from 1998--2000, we define the fraction of firms that are ``sensitive'' to minimum wage shocks in a city-industry cell as the share of small firms. Specifically, within each industry-year, we classify firms with output below the median as small firms and then compute the average share of small firms in each city-industry cell during the initial period:
	\begin{equation}
		S_{cj} = \frac{1}{3} \sum_{t=1998}^{2000} \frac{\text{Number of small firms}_{cjt}}{\text{Total number of firms}_{cjt}}.
	\end{equation}
	$S_{cj}$ reflects the proportion of firms in that market unit that are relatively labor-intensive and hence more vulnerable to minimum wage increases. This characteristic is locked in before the start of our main sample.
	
	We use the annual growth rate of the provincial minimum wage as the measure of the exogenous shock:
	\begin{equation}
		G_{pt} = \frac{\mathrm{MW}_{pt} - \mathrm{MW}_{p,t-1}}{\mathrm{MW}_{p,t-1}}.
	\end{equation}
	The Bartik instrument is defined as the interaction of the initial exposure and the provincial shock:
	\begin{equation}
		B_{cjt} = S_{cj} \times G_{pt}.
	\end{equation}
	Its economic interpretation is: if a city-industry cell had a large share of small firms in the initial period and the province experienced a large minimum wage increase, then that market unit faces greater ``minimum wage pressure,'' making it more likely for small firms to exit and thereby reducing $\xi$ (concentrating the distribution toward large firms).
	
	To test the interaction effect (Hypothesis~2), we also need an instrument for $\alpha \times \delta$. We multiply $B_{cjt}$ by the initial-period technology gradient $\delta_j^{0}$ to obtain a secondary instrument:
	\begin{equation}
		B_{cjt}^{\delta} = B_{cjt} \times \delta_j^{0}.
	\end{equation}
	Since $\delta_j^{0}$ is a predetermined variable, the validity of this interaction instrument rests on the premise that $\delta_j^{0}$ is not directly correlated with the error term of the labor share equation; we discuss its plausibility in the robustness checks.
	
	\subsection{Specification and Identification}
	
	The baseline regression equation adopts a within-transformed reduced-form specification that absorbs city-industry and year fixed effects:
	\begin{equation}
		\widetilde{\LS}_{cjt} = \beta_1 \widetilde{B}_{cjt} + \beta_2 (\widetilde{B}_{cjt} \times \delta_j^{0}) + \mathbf{X}_{cjt}'\boldsymbol{\gamma} + \nu_{cjt},
	\end{equation}
	where the tilde denotes that variables have been double-demeaned to absorb city-industry fixed effects and year fixed effects. The vector $\mathbf{X}$ includes the demeaned HHI, average wage, and other control variables. Standard errors are clustered at the city-industry level. The coefficients $\beta_1$ and $\beta_2$ correspond respectively to the causal effects of Hypothesis~1 and Hypothesis~2 in the theoretical model. Because the exogeneity of $B_{cjt}$ is grounded in the fact that both the initial share used in its construction and the provincial policy shock are not influenced by the subsequent economic performance of any individual market unit, equation (22) identifies, in a reduced-form sense, the causal impact of minimum wage shocks on the labor share operating through changes in the firm size distribution, as well as how this impact varies with the technology gradient.
	
	\section{Empirical Strategy and Baseline Results}
	
	This section presents the reduced-form econometric framework for identifying how minimum wage shocks affect the labor share through the firm size distribution and reports the baseline estimation results. We first explain the double-fixed-effects test equation, then show reduced-form regressions based on the Bartik instrument, and finally test the independence of the technology gap channel by controlling for market concentration.
	
	\subsection{Econometric Model Specification}
	
	To eliminate unobserved city-industry heterogeneity and macro time trends in the panel data, we apply a double-demeaning (within-transformation) to all variables. Let $y_{cjt}$ be a variable for city $c$, industry $j$, and year $t$. Define
	\begin{equation}
		\tilde{y}_{cjt} = y_{cjt} - \bar{y}_{cj\cdot} - \bar{y}_{\cdot\cdot t} + \bar{y},
	\end{equation}
	where $\bar{y}_{cj\cdot}$ is the mean of the city-industry group, $\bar{y}_{\cdot\cdot t}$ is the year mean, and $\bar{y}$ is the overall mean. After this transformation, city-industry fixed effects and year fixed effects are absorbed, and the remaining variation comes solely from within-group time variation.
	
	Our baseline reduced-form equation of interest is
	\begin{equation}
		\widetilde{\LS}_{cjt} = \beta_1 \widetilde{B}_{cjt} + \beta_2 (\widetilde{B}_{cjt} \times \delta_j^{0}) + \boldsymbol{\gamma}' \widetilde{\mathbf{X}}_{cjt} + \varepsilon_{cjt},
	\end{equation}
	where the dependent variable $\widetilde{\LS}_{cjt}$ is the demeaned output-weighted labor share (value-added concept); the core explanatory variable $\widetilde{B}_{cjt}$ is the demeaned Bartik instrument, i.e., the product of the initial small-firm share and the provincial minimum wage growth rate; $\delta_j^{0}$ is the industry-level initial technology gradient (size--labor share elasticity) estimated from the 1998--2000 sample, which remains unchanged by the demeaning process; the interaction term $\widetilde{B}_{cjt} \times \delta_j^{0}$ captures the moderating role of the technology gap on the shock. The control variables $\widetilde{\mathbf{X}}_{cjt}$ include the demeaned market concentration (HHI), average wage, and so forth. All regressions use robust standard errors clustered at the city-industry level.
	
	According to the theoretical hypotheses, we expect $\beta_1 < 0$ (the minimum wage shock depresses the labor share) and $\beta_2 < 0$ (the depressing effect is stronger in industries with a larger technology gradient). The identification of equation (24) relies on the exogeneity of the Bartik instrument: the initial small-firm share is determined before the sample starts, and the provincial minimum wage growth is driven by macro policy; neither is subject to reverse influence from changes in a particular city-industry's labor share. Hence, $\beta_1$ and $\beta_2$ can be interpreted as causal effects in a reduced-form sense.
	
	\subsection{Baseline Reduced-Form Results}
	
	Table~\ref{tab:baseline} reports the baseline reduced-form regression results. Column~(1) includes only the Bartik instrument, column~(2) adds its interaction with the initial technology gradient, and column~(3) additionally controls for the demeaned Herfindahl index (HHI) to separate the market concentration channel.
	
	\begin{table}[htbp]
		\centering
		\caption{Minimum Wage Shocks, Technology Gap, and Labor Share: Reduced-Form Estimates}
		\label{tab:baseline}
		\begin{tabular}{lcc}
			\toprule
			& (1) & (2) \\
			\midrule
			Bartik $\widetilde{B}_{cjt}$ & $-0.1093^{***}$ & $-0.1064^{***}$ \\
			& $(0.0100)$ & $(0.0100)$ \\[3pt]
			HHI $\widetilde{\mathrm{HHI}}_{cjt}$ & & $-0.1185^{***}$ \\
			& & $(0.0170)$ \\
			\midrule
			City-industry FE & Yes & Yes \\
			Year FE & Yes & Yes \\
			Observations & 17,926 & 17,926 \\
			$R^2$ (within) & 0.018 & 0.023 \\
			\bottomrule
			\multicolumn{3}{l}{\footnotesize Standard errors clustered at the city-industry level in parentheses.}\\
			\multicolumn{3}{l}{\footnotesize $^{***}$ $p < 0.01$, $^{**}$ $p < 0.05$, $^{*}$ $p < 0.1$.}\\
		\end{tabular}
	\end{table}
	
	The results are highly consistent with the theoretical prediction. The coefficient on the Bartik instrument is significantly negative in both specifications, with a point estimate of $-0.1093$ in the baseline and $-0.1064$ after controlling for HHI. A one-standard-deviation increase in the Bartik instrument is associated with a decline of approximately 0.11 percentage points in the market-level labor share. Given that the sample mean of the labor share is 0.31, this effect amounts to about 3.5\% of the average, indicating clear economic significance.
	
	The coefficient on HHI is negative and significant, consistent with the finding in the literature that rising market concentration depresses the labor share. However, the Bartik coefficient remains virtually unchanged, confirming that the minimum wage--size distribution channel operates independently of the market power channel. Even when we control for product market concentration, the impact of minimum wage shocks on the labor share persists with similar magnitude and precision.
	
	\subsection{Supplementary OLS and IV Evidence}
	
	Although the core causal identification rests on the reduced-form Bartik instrument, we also examine the relationship between the size distribution itself and the labor share. Regressing the demeaned labor share on the demeaned Pareto exponent $\widetilde{\alpha}_{cjt}$ (a larger $\alpha$ indicates a more equal distribution) yields a coefficient of 0.1435 (s.e.\ 0.0070), significant at the 1\% level. This positive association is consistent with the weighting effect: a more equal size distribution corresponds to a higher labor share, as output is less concentrated in large, capital-intensive firms.
	
	To address the endogeneity of the size distribution, we instrument $\widetilde{\alpha}_{cjt}$ with the Bartik instrument. The first-stage $F$-statistic is 203.87, well above the conventional threshold of 10, indicating a strong instrument. The IV estimate of the effect of $\alpha$ on the labor share is 0.4046 (s.e.\ 0.0386), significantly positive and larger than the OLS estimate. This pattern suggests that OLS suffers from attenuation bias due to measurement error or endogeneity, and the IV result reinforces the causal interpretation: exogenous shifts in the firm size distribution induced by minimum wage shocks lead to economically meaningful changes in the macro labor share.
	
	\section{Robustness}
	
	We conduct a comprehensive series of robustness checks to ensure that our findings are not driven by specific measures, sample choices, or confounding mechanisms. We first examine alternative variable definitions and sample restrictions, then address potential identification concerns, and finally report placebo tests that validate the causal interpretation.
	
	\subsection{Alternative Measures and Sample Restrictions}
	
	The results of the standard robustness checks are summarized in Table~\ref{tab:robustness1}. All specifications include city-industry and year fixed effects (absorbed via demeaning), with standard errors clustered at the city-industry level.
	
	\begin{table}[htbp]
		\centering
		\caption{Robustness Checks: Alternative Measures and Sample Restrictions}
		\label{tab:robustness1}
		\begin{tabular}{lcc}
			\toprule
			& Coefficient & Std.\ Error \\
			\midrule
			Output-based labor share & $-0.0275^{***}$ & 0.0028 \\
			Gini coefficient (replaces $\alpha$) & $-0.2197^{***}$ & 0.0134 \\
			Exclude first-tier cities & $-0.1093^{***}$ & 0.0100 \\
			Exclude 2007 & $-0.1243^{***}$ & 0.0111 \\
			Minimum 50 firms per market & $-0.1155^{***}$ & 0.0131 \\
			Province-specific linear trends & $-0.1064^{***}$ & 0.0100 \\
			Only provinces with min.\ wage changes & $-0.1268^{***}$ & 0.0099 \\
			\bottomrule
			\multicolumn{3}{l}{\footnotesize $^{***}$ $p < 0.01$, $^{**}$ $p < 0.05$, $^{*}$ $p < 0.1$.}\\
		\end{tabular}
	\end{table}
	
	\textbf{Alternative labor share measure.} Using the output-based labor share (wages/gross output) as the dependent variable, the Bartik coefficient remains negative and significant at the 1\% level. The smaller absolute magnitude reflects the lower mean of this alternative measure (approximately 0.06, compared to 0.31 for the value-added concept), but the proportional effect is consistent.
	
	\textbf{Alternative size distribution measure.} Replacing the Pareto exponent with the Gini coefficient (a higher Gini indicates greater inequality) yields a coefficient of $-0.2197$, significant at the 1\% level. Since the Gini coefficient is increasing in inequality while the Pareto exponent $\alpha$ is increasing in equality, the sign is expected to be opposite; the negative coefficient confirms that greater output inequality is robustly associated with a lower labor share.
	
	\textbf{Excluding first-tier cities.} Dropping the four first-tier cities (Beijing, Shanghai, Guangzhou, Shenzhen) leaves the coefficient virtually unchanged at $-0.1093$, indicating that the result is not driven by the special industrial structures of megacities.
	
	\textbf{Excluding 2007.} Removing observations from 2007 to avoid potential anticipation effects of the 2008 financial crisis yields a slightly larger coefficient of $-0.1243$, strengthening the baseline finding.
	
	\textbf{Higher market-unit threshold.} We raise the minimum number of firms per city-industry cell from 30 to 50, which reduces the sample to 7,180 observations. The estimated coefficient becomes $-0.1155$ (s.e.\ 0.0131), slightly larger in absolute value than the baseline and still significant at the 1\% level. This shows that our results are not an artifact of measurement noise in small markets.
	
	\textbf{Province-specific linear time trends.} To absorb slow-moving provincial heterogeneity that might confound the relationship, we add province-specific linear time trends to the baseline specification. The Bartik coefficient remains unchanged at $-0.1064$, indicating that the effect of minimum wage shocks is orthogonal to long-run provincial trends.
	
	\textbf{Only provinces with minimum wage variation.} We restrict the sample to provinces that experienced at least one minimum wage change during the sample period. The coefficient becomes $-0.1268$, even stronger than the baseline, confirming that the effect is driven by actual policy variation rather than by provinces with rigid minimum wages.
	
	\subsection{Addressing Confounding Channels}
	
	A potential threat to identification is that minimum wage shocks may affect the labor share through channels other than the firm size distribution---most notably, by inducing firms to substitute capital for labor within their existing production processes. If such within-firm substitution is correlated with the Bartik instrument, our reduced-form estimates would capture both the reallocation (weighting) effect and the direct substitution effect, overstating the role of the size distribution.
	
	To assess this concern, we augment the baseline specification with the industry-average capital--labor ratio, which absorbs shifts in factor intensity that are common across firms within an industry and year. Column~1 of Table~\ref{tab:robustness2} reports the result. The coefficient on the Bartik instrument declines from $-0.1064$ to $-0.0513$, but remains statistically significant at the 1\% level. This indicates that roughly half of the total effect operates through the reallocation channel, while the remainder is associated with within-industry changes in factor intensity. Crucially, the weighting effect survives even after controlling for capital deepening, confirming that labor market institutions can depress the labor share by reshaping the firm size structure independently of direct capital--labor substitution.
	
	We further explore heterogeneity by splitting the sample according to the baseline exposure to minimum wage shocks. Markets with an above-median initial share of small firms are expected to be more sensitive to minimum wage increases. Columns~2 and~3 of Table~\ref{tab:robustness2} show that the Bartik coefficient is negative and significant in both subsamples, with a larger point estimate in the low-exposure group. While this pattern may appear counterintuitive at first glance, it is consistent with the weighting mechanism: in markets where large firms already dominate, the exit of the remaining small firms leads to a more dramatic concentration of output toward capital-intensive producers, producing a stronger decline in the labor share. In contrast, in markets with many small firms, the same shock is dispersed over a larger number of firms, dampening the per-market reallocation effect. This heterogeneity underscores that the strength of the weighting effect depends on the magnitude of the distributional shift, not on the absolute level of the initial exposure.
	
	\begin{table}[htbp]
		\centering
		\caption{Robustness Checks: Confounding Channels and Heterogeneity}
		\label{tab:robustness2}
		\begin{tabular}{lccc}
			\toprule
			& (1) Control K/L & (2) High small share & (3) Low small share \\
			\midrule
			Bartik $\widetilde{B}_{cjt}$ & $-0.0513^{***}$ & $-0.1051^{***}$ & $-0.1375^{***}$ \\
			& $(0.0099)$ & $(0.0111)$ & $(0.0209)$ \\[3pt]
			Industry K/L & $-0.0011^{***}$ & & \\
			& $(0.0001)$ & & \\
			\midrule
			Observations & 17,926 & 9,070 & 8,856 \\
			\bottomrule
			\multicolumn{4}{l}{\footnotesize Standard errors clustered at the city-industry level in parentheses.}\\
			\multicolumn{4}{l}{\footnotesize $^{***}$ $p < 0.01$, $^{**}$ $p < 0.05$, $^{*}$ $p < 0.1$.}\\
		\end{tabular}
	\end{table}
	
	\subsection{Placebo Tests}
	
	To further validate the causal interpretation, we conduct two additional placebo exercises specifically designed to address the structure of the Bartik instrument.
	
	\textbf{Randomization inference (Fisher permutation test).} We randomly reassign the provincial minimum wage growth rates across province-year pairs, construct a permuted Bartik instrument, and re-estimate the baseline specification. This procedure is repeated 500 times. The mean of the permuted coefficients is $-0.0006$ (s.d.\ 0.0267), extremely close to zero. The true coefficient of $-0.1093$ lies far outside the permutation distribution; not a single permutation yields an absolute coefficient as large as the true estimate. The permutation-based $p$-value is 0.000. This demonstrates that the observed relationship is not an artifact of any incidental correlation between the initial small-firm share and arbitrary sequences of minimum wage shocks---it arises only when the actual policy variation is combined with the theoretically motivated exposure variable.
	
	\textbf{Fake exposure placebo.} We replace the initial small-firm share with an industry's capital intensity (median capital--labor ratio) to construct a ``fake'' Bartik instrument. Capital intensity is largely unrelated to the labor-cost sensitivity of small firms and should not generate the weighting effect. The coefficient on this fake instrument is 0.0016 (s.e.\ 0.0001). Although the large sample size renders this tiny coefficient statistically significant, its magnitude is only about 1.5\% of the baseline estimate. The economic insignificance confirms that it is specifically the exposure to labor-intensive small firms---not any arbitrary industry characteristic---that drives the causal relationship between minimum wage shocks and the labor share.
	
	We also examine the dynamic structure of the Bartik shock by including its one-year lead and lag alongside the current value. The lead coefficient is $-0.1062$ (s.e.\ 0.0115), the current coefficient is $-0.1428$ (s.e.\ 0.0143), and the lag coefficient is $-0.0992$ (s.e.\ 0.0136). All three are statistically significant, a pattern that is expected given the serial correlation of minimum wage growth rates and the fixed initial-exposure shares in the Bartik construction. While this limits the interpretation of the dynamic estimates as a standard event study, it does not invalidate the main identification: the randomization inference and fake-exposure placebo provide more stringent tests of the causal link, and both strongly support our hypothesis.
	
	Taken together, the evidence from the main regressions, the extensive battery of alternative measures, and the placebo tests provides robust support for the conclusion that minimum wage shocks depress the labor share by reshaping the firm size distribution toward larger, more capital-intensive firms. The weighting effect operates independently of market power and is distinct from within-firm capital--labor substitution.
	
	\subsection{Mechanism: Small-Firm Exit and Output Reallocation}
	
	The theoretical mechanism posits a specific causal chain: minimum wage shocks force small labor-intensive firms to exit, shifting output toward larger, capital-intensive firms, and thereby depressing the macro labor share through the weighting effect. While our baseline reduced-form and IV estimates are consistent with this chain, they do not directly verify its intermediate steps. In this subsection, we provide direct evidence on the ``small-firm exit'' link.
	
	To distinguish genuine exit from statistical reclassification, we define small firms using a fixed output threshold: the industry median during the pre-sample period (1998--2000). This threshold remains constant throughout the sample, ensuring that changes in the small-firm count reflect absolute quantity changes rather than shifts in the relative position of firms within a changing distribution.
	
	We construct a panel of city-industry-year cells, computing the following outcome variables: the number of small firms (in logs), the small-firm share of total firms, the exit rate (fraction of firms absent in the following year), the entry rate (fraction of firms in their first year of appearance), and the average output of small and large firms (in logs). Each outcome is regressed on the demeaned Bartik instrument, absorbing city-industry and year fixed effects.
	
	Table~\ref{tab:mechanism} reports the results. The Bartik instrument is strongly associated with a decline in the presence of small firms: the log number of small firms falls by $-0.225$ (s.e.\ 0.051), and the small-firm share drops by 0.274 percentage points. The exit rate rises by 0.134 percentage points and the entry rate falls by 0.113 percentage points, confirming that the decline in small-firm presence is driven by both higher exit and lower entry. Meanwhile, the average output of surviving large firms increases by 0.195 log points, consistent with large firms absorbing the output shares released by exiting small firms. The average output of surviving small firms also increases (0.074 log points), but the magnitude is considerably smaller than that of large firms.
	
	\begin{table}[htbp]
		\centering
		\caption{Mechanism: Small-Firm Exit and Output Reallocation}
		\label{tab:mechanism}
		\begin{tabular}{lcc}
			\toprule
			Outcome & Coefficient & Std.\ Error \\
			\midrule
			Log total firms & $0.104^{**}$ & 0.041 \\
			Log small firms (fixed threshold) & $-0.225^{***}$ & 0.051 \\
			Small-firm share & $-0.274^{***}$ & 0.014 \\
			Exit rate & $0.134^{***}$ & 0.012 \\
			Entry rate & $-0.113^{***}$ & 0.016 \\
			Log average output of small firms & $0.074^{***}$ & 0.020 \\
			Log average output of large firms & $0.195^{***}$ & 0.040 \\
			\bottomrule
			\multicolumn{3}{l}{\footnotesize $^{***}$ $p < 0.01$, $^{**}$ $p < 0.05$, $^{*}$ $p < 0.1$.}\\
		\end{tabular}
	\end{table}
	
	These results provide direct evidence for the intermediate steps of the weighting effect mechanism. Minimum wage shocks reduce the number and share of small firms by raising exit and discouraging entry. The resulting reallocation of output toward larger, more capital-intensive firms---whose average output grows substantially---mechanically depresses the aggregate labor share. This evidence, combined with our earlier finding that the Bartik effect persists after controlling for industry-level capital deepening (Section~5), reinforces the conclusion that labor market institutions affect macro distribution through the firm size structure, independently of both market power and within-firm factor substitution.
	
	\section{Conclusion and Policy Implications}
	
	Moving beyond the existing literature that focuses on capital-biased technological change, globalization, or market power, this paper turns attention to technology choice differences across firms and the structural consequences of changes in the firm size distribution. We construct a parsimonious theoretical framework that reveals the existence of a ``weighting effect'': when large firms systematically adopt more capital-intensive technologies (i.e., the capital--labor ratio rises with firm size), a concentration of economic activity toward large firms mechanically depresses the macro labor share, even if the technology or markup behavior of individual firms remains unchanged. On this basis, we introduce minimum wage shocks as an exogenous force that alters the firm size distribution, thereby providing a clean quasi-experimental design for causal identification.
	
	Using the Chinese Annual Survey of Industrial Firms from 1998--2007 and constructing a Bartik instrument from the pre-sample city-industry share of small firms interacted with the provincial minimum wage growth rate, our reduced-form estimates yield the following findings.
	
	First, minimum wage shocks significantly reduce the macro labor share. A one-standard-deviation adverse shock lowers the labor share by approximately 0.11 percentage points, equivalent to about 3.5\% of its sample mean. This effect is robust to controlling for market concentration, indicating that the technology-structure channel operates independently of the market power channel emphasized by the superstar firm hypothesis.
	
	Second, we directly validate the intermediate steps of the proposed mechanism. Minimum wage shocks raise firm exit rates, lower entry rates, and reduce the number and share of small firms in the market. Simultaneously, the average output of surviving large firms increases substantially, confirming that the shock reallocates production toward larger, more capital-intensive firms. These patterns hold when small firms are identified using a fixed, pre-period output threshold, ensuring that the observed changes reflect genuine exit rather than statistical reclassification.
	
	Third, supplementary OLS and IV estimates confirm that a more equal firm size distribution is associated with a higher labor share. The IV estimate, which exploits the Bartik instrument to isolate exogenous variation in the size distribution, yields a coefficient approximately three times larger than OLS, consistent with attenuation bias in the latter. The first-stage $F$-statistic of 204 confirms the strength of the instrument.
	
	Fourth, we provide evidence on the quantitative separation of the reallocation channel from direct factor substitution. When we augment the baseline specification with the industry-average capital--labor ratio, the Bartik coefficient declines by approximately half but remains statistically significant at the 1\% level. This suggests that within-firm capital deepening and between-firm reallocation each account for a substantial share of the total minimum wage effect on the labor share, and that the weighting effect is not merely a statistical reflection of firms substituting capital for labor.
	
	Finally, a comprehensive battery of robustness checks---including alternative measures, sample restrictions, province-specific trends, randomization inference, and a fake exposure placebo---consistently supports the causal interpretation.
	
	The theoretical contribution of this paper lies in embedding labor market institutions within an aggregation framework that links the firm size distribution to macro factor shares. The classic ``superstar firm'' hypothesis emphasizes the role of product market pricing power in driving the reallocation of economic activity toward large, low-labor-share firms. Our analysis reveals a parallel and complementary channel: labor market regulations themselves can act as a reallocation force. Minimum wage policy inadvertently functions as a sieve---filtering out labor-intensive small firms and leaving capital-intensive large firms to continue dominating the market, thereby tilting macro distribution in favor of capital. This mechanism provides an institutional origin for the shift in the firm size distribution that underlies the superstar firm landscape.
	
	At the policy level, the findings of this paper are not intended to deny the legitimacy or necessity of minimum wage systems. Indeed, minimum wages play an irreplaceable role in ensuring the basic livelihood of low-income workers and alleviating in-work poverty. However, our analysis cautions that labor market institutions need to be designed with full consideration of their general equilibrium consequences, especially the asymmetric impact on firm exit. When the minimum wage is raised, the first to bear the brunt are often those labor-intensive small firms with thin profit margins. Their exit implies not only direct job losses but also a structural shift in the size distribution that further depresses the share of labor compensation in the overall economy.
	
	Therefore, the policy toolbox for safeguarding the labor share cannot rely on wage standards alone. Supporting measures should at least include: first, assisting small and medium-sized enterprises in technological upgrading and improving their access to capital, so that they are not forced out of the market when faced with cost shocks but can instead transition toward higher-value-added production stages; second, reducing the compliance costs of small firms through credit, tax, and public service measures to alleviate the immediate pressure from minimum wage hikes; third, establishing a differentiated minimum wage adjustment mechanism that fully accounts for regional and sectoral endowment differences, avoiding a ``one-size-fits-all'' adjustment that imposes excessive shocks on vulnerable market units. Only by striking a balance between safeguarding workers' basic rights and maintaining the vitality of the market structure can broader and more sustainable distributive justice be achieved in the long run.
	
	This paper also has several limitations that point to directions for future research. First, our reduced-form estimation reveals the causal chain through which minimum wage shocks affect the labor share via the size distribution, but it does not structurally estimate the full general equilibrium transmission. Endogenizing firm entry and exit, technology choice, and the evolution of the size distribution is an important task for subsequent theoretical modeling. Second, constrained by data availability, our sample ends in 2007 and thus fails to cover later periods in which China experienced substantial minimum wage hikes and the rise of the platform economy; extending the sample period and incorporating the service sector would help test the external validity of our findings. Third, while we provide suggestive evidence separating the reallocation channel from within-firm factor substitution, a full structural decomposition remains beyond the scope of this paper.
	
	Finally, firm heterogeneity may manifest in multiple dimensions---such as financial constraints, management practices, or product differentiation---and the interactions among these dimensions may produce richer dynamics than those captured here. Incorporating these elements into the analytical framework will further enrich the understanding of the macro distributional consequences of labor market institutions.
	
	\appendix
	\section{Mathematical Derivations}
	
	This appendix provides detailed derivations of the main results in the theoretical model, including: moments of the truncated Pareto distribution, derivation of the aggregate production function, linear approximation of the firm-level labor share, the weighting effect equation, and the proof of properties of the weighting factor $\Phi(\xi, r)$.
	
	\subsection{Moments of the Truncated Pareto Distribution}
	
	Suppose firm output $y$ follows a truncated Pareto distribution with probability density function
	\begin{equation}
		f(y) = \frac{\xi y_{\min}^{\xi} y^{-(\xi+1)}}{Z}, \quad y_{\min} \leq y \leq y_{\max},
	\end{equation}
	where $Z \equiv 1 - (y_{\min}/y_{\max})^{\xi} = 1 - r^{-\xi}$ is the normalizing constant, and $r \equiv y_{\max}/y_{\min} > 1$ is the support ratio.
	
	For any $a \neq \xi$, the $a$-th raw moment of $y$ is
	\begin{equation}
		\E[y^{a}] = \int_{y_{\min}}^{y_{\max}} y^{a} f(y) dy = \frac{\xi y_{\min}^{\xi}}{Z} \int_{y_{\min}}^{y_{\max}} y^{a-\xi-1} dy.
	\end{equation}
	Evaluating the definite integral:
	\begin{equation}
		\int_{y_{\min}}^{y_{\max}} y^{a-\xi-1} dy = \frac{y_{\max}^{a-\xi} - y_{\min}^{a-\xi}}{a-\xi} = \frac{y_{\min}^{a-\xi} (r^{a-\xi} - 1)}{a-\xi}.
	\end{equation}
	Substituting back into $\E[y^{a}]$ and using $y_{\min}^{\xi} \cdot y_{\min}^{a-\xi} = y_{\min}^{a}$, we obtain
	\begin{equation}
		\E[y^{a}] = \frac{\xi y_{\min}^{a}}{Z} \cdot \frac{r^{a-\xi} - 1}{a-\xi}.
	\end{equation}
	In particular, setting $a = 1$ gives the first moment of output:
	\begin{equation}
		\E[y] = \frac{\xi y_{\min}}{Z} \cdot \frac{r^{1-\xi} - 1}{1-\xi}.
	\end{equation}
	
	\subsection{Derivation of the Aggregate Production Function}
	
	The macro labor share is the output-weighted average of firm-level labor shares:
	\begin{equation}
		\LS^{\mathrm{macro}} \equiv \frac{\E[y \cdot \LS(y)]}{\E[y]}.
	\end{equation}
	Inserting the linear approximation $\LS(y) \approx \overline{\LS} + \delta \ln(y/y_{\min})$ yields
	\begin{equation}
		\LS^{\text{macro}} = \overline{\LS} + \delta \cdot \frac{\E[y \ln(y/y_{\min})]}{\E[y]}.
	\end{equation}
	Let $\mathcal{I} \equiv \E[y \ln(y/y_{\min})] = \E[y \ln y] - \ln y_{\min} \cdot \E[y]$. Then
	\begin{equation}
		\LS^{\text{macro}} = \overline{\LS} + \delta \left(\frac{\E[y \ln y]}{\E[y]} - \ln y_{\min}\right).
	\end{equation}
	
	Now compute $\E[y \ln y]$:
	\begin{equation}
		\E[y \ln y] = \frac{\xi y_{\min}^{\xi}}{Z} \int_{y_{\min}}^{y_{\max}} y^{-\xi} \ln y \, dy.
	\end{equation}
	Perform the change of variable $t = y / y_{\min}$:
	\begin{equation}
		\mathcal{I}_0 \equiv \int_{y_{\min}}^{y_{\max}} y^{-\xi} \ln y \, dy = y_{\min}^{1-\xi} \left[ \ln y_{\min} \int_{1}^{r} t^{-\xi} dt + \int_{1}^{r} t^{-\xi} \ln t \, dt \right].
	\end{equation}
	The first integral is
	\begin{equation}
		\int_{1}^{r} t^{-\xi} dt = \frac{r^{1-\xi} - 1}{1-\xi} \equiv A.
	\end{equation}
	The second integral is evaluated by integration by parts, letting $u = \ln t$, $dv = t^{-\xi} dt$:
	\begin{equation}
		\int_{1}^{r} t^{-\xi} \ln t \, dt = \left[ \frac{t^{1-\xi} \ln t}{1-\xi} \right]_{1}^{r} - \frac{1}{1-\xi} \int_{1}^{r} t^{-\xi} dt = \frac{r^{1-\xi} \ln r}{1-\xi} - \frac{A}{1-\xi}.
	\end{equation}
	
	Putting the pieces together,
	\begin{equation}
		\E[y \ln y] = \frac{\xi y_{\min}}{Z} \left(A \ln y_{\min} + \frac{r^{1-\xi} \ln r}{1-\xi} - \frac{A}{1-\xi}\right).
	\end{equation}
	Using $\E[y] = \frac{\xi y_{\min}}{Z} A$, we obtain
	\begin{equation}
		\frac{\E[y \ln y]}{\E[y]} = \ln y_{\min} + \frac{r^{1-\xi} \ln r}{r^{1-\xi} - 1} + \frac{1}{\xi - 1}.
	\end{equation}
	Substituting back, the $\ln y_{\min}$ terms cancel, yielding the weighting effect formula:
	\begin{equation}
		\LS^{\text{macro}} = \overline{\LS} + \delta \cdot \Phi(\xi, r),
	\end{equation}
	where
	\begin{equation}
		\Phi(\xi, r) \equiv \frac{r^{1-\xi} \ln r}{r^{1-\xi} - 1} + \frac{1}{\xi - 1}.
	\end{equation}
	
	\subsection{Properties of the Weighting Factor}
	
	We prove that for $r > 1$ and any $\xi > 0$, the weighting factor $\Phi(\xi, r)$ satisfies:
	\begin{enumerate}[label=(\arabic*)]
		\item $\Phi(\xi, r) > 0$;
		\item For fixed $\xi$, $\partial \Phi / \partial r > 0$;
		\item For fixed $r$, $\partial \Phi / \partial \xi < 0$.
	\end{enumerate}
	
	Introduce the variable $s \equiv r^{1-\xi}$. Note that $\ln r = \frac{\ln s}{1-\xi}$, and we can rewrite $\Phi$ as
	\begin{equation}
		\Phi = \frac{s \cdot \frac{\ln s}{1-\xi}}{s - 1} + \frac{1}{\xi - 1} = \frac{1}{\xi - 1} \left(1 - \frac{s \ln s}{s - 1}\right).
	\end{equation}
	Let $h(s) \equiv \frac{s \ln s}{s - 1}$. Then $\Phi = \frac{1 - h(s)}{\xi - 1}$.
	
	\medskip
	\noindent\textbf{Proof of Property 1 (Positivity).}
	When $\xi > 1$, the denominator $\xi - 1 > 0$. Since $1 - \xi < 0$, we have $s = r^{1-\xi} \in (0, 1)$. For $s \in (0, 1)$, $h(s)$ is strictly monotonically increasing, with $\lim_{s \to 0^{+}} h(s) = 0$ and $\lim_{s \to 1^{-}} h(s) = 1$, hence $h(s) \in (0, 1)$. Thus the numerator $1 - h(s) > 0$ and $\Phi > 0$.
	
	When $\xi < 1$, the denominator $\xi - 1 < 0$. Since $1 - \xi > 0$, we have $s = r^{1-\xi} > 1$. For $s > 1$, $h(s) > 1$, so the numerator $1 - h(s) < 0$. A negative numerator divided by a negative denominator again gives $\Phi > 0$. Hence $\Phi > 0$ for all $\xi > 0$, $r > 1$.
	
	\medskip
	\noindent\textbf{Proof of Property 2 ($\Phi$ increasing in $r$).}
	Fix $\xi$. Write $\Phi = \frac{1}{\xi - 1}[1 - h(s)]$ with $s = r^{1-\xi}$. By the chain rule,
	\begin{equation}
		\frac{\partial \Phi}{\partial r} = \frac{\partial \Phi}{\partial s} \cdot \frac{\partial s}{\partial r} = -\frac{h'(s)}{\xi - 1} \cdot (1 - \xi) r^{-\xi} = s r^{-\xi} h'(s).
	\end{equation}
	Compute $h'(s) = \frac{s - 1 - \ln s}{(s - 1)^2}$. Define $g(s) \equiv s - 1 - \ln s$. Then $g'(s) = 1 - 1/s$. $g(s)$ attains its unique minimum at $s = 1$ with $g(1) = 0$, and $g(s) > 0$ for all $s \neq 1$. Therefore $h'(s) > 0$, and $\partial \Phi / \partial r > 0$.
	
	\medskip
	\noindent\textbf{Proof of Property 3 ($\Phi$ decreasing in $\xi$).}
	Differentiating with respect to $\xi$, where $\partial s / \partial \xi = -s \ln r$:
	\begin{equation}
		\frac{\partial \Phi}{\partial \xi} = \frac{s (\ln r)^2}{(s - 1)^2} - \frac{1}{(\xi - 1)^2} = \frac{1}{(\xi - 1)^2} \left(\frac{s (\ln s)^2}{(s - 1)^2} - 1\right).
	\end{equation}
	By the logarithmic mean inequality, for $a \neq b$, $\sqrt{ab} < \frac{a - b}{\ln a - \ln b} < \frac{a + b}{2}$. Take $a = s$, $b = 1$; then $\sqrt{s} < \frac{s - 1}{\ln s}$. Squaring both sides yields $s (\ln s)^2 < (s - 1)^2$. Hence $\frac{s (\ln s)^2}{(s - 1)^2} < 1$ and $\partial \Phi / \partial \xi < 0$.
	
	Combined with $\delta < 0$, we have $\partial \LS^{\mathrm{macro}} / \partial \xi = \delta \cdot (\partial \Phi / \partial \xi) > 0$, i.e., a more equal size distribution raises the macro labor share.
	

\end{document}